 \documentclass[
preprintnumbers,prd,a4paper,showpacs, amsmath, amssymb, floatfix,
nofootinbib,wrapfloat byrevtex,article,report]{revtex4}

 \usepackage{amsmath,mathrsfs,graphicx,epstopdf,placeins,float}
 \usepackage{booktabs}
 \usepackage{multirow}
\usepackage{bbm,verbatim}
\usepackage{fancyhdr}
\usepackage{lipsum}

\usepackage{lineno,hyperref}
\modulolinenumbers[5]
 \newcommand{\beq}[1]{\begin{equation}\label{#1}}
 \newcommand{\eeq}{\end{equation}}
 \newcommand{\bea}[1]{\begin{eqnarray}\label{#1}}
 \newcommand{\eea}{\end{eqnarray}}
 \newcommand\figcaption{\def\@captype{figure}\caption}
 \newcommand\tabcaption{\def\@captype{table}\caption}

\fancypagestyle{firstpage}{
  \fancyhf{}
  \fancyhead[C]{Submitted to Chinese Physics C}
  \fancyfoot[C]{-~\thepage~-}
}
\begin{document}
\title{Some New Symmetric Relations and the Prediction of Left and Right Handed Neutrino Masses using Koide's Relation }
\author{Yong-Chang Huang, Syeda Tehreem Iqbal, Zhen Lei and Wen-Yu Wang } 
\affiliation{Institute of Theoretical Physics, Beijing University of Technology, Beijing, 100124, China}
\begin{abstract}

Masses of the three generations of charged leptons are known to completely satisfy the
Koide's mass relation. But the question remains if such a relation exists for neutrinos?
In this paper, by considering SeeSaw mechanism as the mechanism generating tiny neutrino masses, we show how neutrinos satisfy the Koide's mass relation, on the basis of which we systematically give exact values of not only left but also right handed neutrino masses.
\end{abstract}
\maketitle
\thispagestyle{firstpage}

%

\section{Introduction}

Despite being the most successful model of particle physics,
Standard Model (SM) fails to answer many questions like why the parameters of SM
are the way they are? Is there any relation among these parameters?
Why the Koide's relation for the charged lepton is 2/3? Yoshio Koide  \cite{Koide,de} pointed out that a very simple relationship exists for the pole masses (given in the Table {\ref{I}}) of the three generations of charged leptons,
\begin{equation}
(m_e+m_\mu+m_\tau) = \frac {2}{3} (\sqrt{m_e}+\sqrt{m_\mu}+\sqrt{m_\tau})^2.
\label{koideee}
\end{equation}
which is surprisingly precise to a good degree of accuracy.
This precision inspired Koide to propose models \cite{k1, k2, k22} in an attempt to explain the underlying physics.
Various attempts have been made to extend this formula to other particles. In \cite{Strange} some speculations related to the extension of Eq. [\ref{koideee}] to quark and leptons are given along with its relations to recent theoretical developments.
Different ideas following the implementation of this formula can also be found \cite{JM,Ma,Wr,implement}.
A geometric interpretation for Koide's relation was given in \cite{Foot} in which the square root of the mass of leptons
$\sqrt{m_{_{l}}}$ is used to construct a vector $\overrightarrow{V}$, such that
\begin{equation}
\overrightarrow{V}=\left( \sqrt{m_{e}}, \quad \sqrt{m_{\mu}}, \quad \sqrt{m_{\tau}}\right),
\end{equation}
then Koide's formula can be considered equivalent to the angle between the vector $\left(1,1,1\right)$
and $\overrightarrow{V}$ which is $\frac{\pi}{4}$, which will be considered, in details, in Sect. III. The questions that follow from the above interpretation are: why the vector is $\left(1,1,1\right)$
and why is the angle $\frac{\pi}{4}$ ? The aim of this paper is to give a meaning to the geometric interpretation
and to extend Koide's formula to neutrinos such that the masses of left handed and right handed neutrinos can be predicted.

\begin{center}
Table {\ref{I}}: Mass of leptons
\end{center}
\begin{center}
\begin{tabular}{|c|c|}
\hline
\quad lepton \quad &  mass \quad \quad (MeV)\\    \hline
$e$      &  $0.510998928\pm0.000000011$ \\ \hline
$\mu$    &  $105.6583715\pm0.0000035$ \\ \hline
$\tau$   &  $1776.82\pm0.16$ \\ \hline
\end{tabular}\label{I}
\end{center}

The plan of the paper is as follow: In Sect.II we find two analytical formulas to achieve the masses of neutrinos using the data provided by experiments. Sect.III is served to give a meaning to the geometrical interpretation given by Foot\cite{Foot}. In Sect.IV we devise a formula to find the value of right handed neutrino mass terms. Considering a relation between left and right handed neutrinos we can solve the analytical formulas for neutrino masses, details of which are given in Sect.V. The last section is the summary and conclusion.


\section{Analytical Formula For Neutrino Masses}
The neutrino mass term has the form
\begin{equation}
\label{3}
\mathcal{L}=\frac{1}{2}
\left(\begin{array}{cc}
\bar{\nu}_{l} & \bar{\nu}_{R}^{c}
\end{array}\right)
\mathcal{M}
\left(\begin{array}{c}
\nu_{l}^{c}\\
\nu_{R}
\end{array}\right)
+h.c. \quad
\end{equation}
\indent Supposing that the mass matrix
$\mathcal{M}_{mass}$ can be diagonalized as follow
\begin{equation}
\left(\begin{array}{cccccc}
0 & 0 & 0 & m_{\scriptsize \texttt{D}_{1}} & 0 & 0\\
0 & 0 & 0 & 0 & m_{\scriptsize \texttt{D}_{2}} & 0\\
0 & 0 & 0 & 0 & 0 & m_{\scriptsize \texttt{D}_{3}}\\
m_{\scriptsize \texttt{D}_{1}} & 0 & 0 & M_{1} & 0 & 0\\
0 & m_{\scriptsize \texttt{D}_{2}} & 0 & 0 & M_{2} & 0\\
0 & 0 & m_{\scriptsize \texttt{D}_{3}} & 0 & 0 & M_{3}\\
\end{array}\right)\\
\label{matrix}
\end{equation}
where $M_{1}$, $ M_{2}$ and $M_{3}$ are the Majorana mass
coefficients.
\\
\indent The eigenvalues of the matrix are:
\begin{eqnarray}
&~&\frac{1}{2}M_{1}\pm \frac{1}{2}\sqrt{M_{1}^2+4m_{\scriptsize \texttt{D}_{1}}^2} ,\nonumber \\
&~&\frac{1}{2}M_{2}\pm \frac{1}{2}\sqrt{M_{2}^2+4m_{\scriptsize \texttt{D}_{2}}^2} ,\nonumber \\
&~&\frac{1}{2}M_{3}\pm \frac{1}{2}\sqrt{M_{3}^2+4m_{\scriptsize \texttt{D}_{3}}^2} .
\label{eigenvalue1}
\end{eqnarray}
When $M_{i} \gg m_{\scriptsize \texttt{D}_{i}}  (i=1, 2, 3)$, the neutrino
masses would be
\begin{equation}
\frac{m_{\scriptsize \texttt{D}_{1}}^2}{M_{1}}, \quad
 \frac{m_{\scriptsize \texttt{D}_{2}}^2}{M_{2}}, \quad
 \frac{m_{\scriptsize \texttt{D}_{3}}^2}{M_{3}},
\label{6}
\end{equation}
which is just the SeeSaw mechanism. The strict form of
$\mathcal{M}_{mass}$ is given by
\begin{equation}
\left(\begin{array}{cccccc}
0 & 0 & 0 & m_{\scriptsize \texttt{D}_{1}} & 0 & 0\\
0 & 0 & 0 & 0 & m_{\scriptsize \texttt{D}_{2}} & 0\\
0 & 0 & 0 & 0 & 0 & m_{\scriptsize \texttt{D}_{3}}\\
m_{\scriptsize \texttt{D}_{1}} & 0 & 0 & M_{1} & m_{_{1}} & m_{_{2}}\\
0 & m_{\scriptsize \texttt{D}_{2}} & 0 & m_{_{1}} & M_{2} & m_{_{3}}\\
0 & 0 & m_{\scriptsize \texttt{D}_{3}} & m_{_{2}} & m_{_{3}}& M_{3}\\
\end{array}\right) . \\
\label{mat}
\end{equation}
$m_{\scriptsize \texttt{D}_{1}},m_{\scriptsize \texttt{D}_{2}},
 m_{\scriptsize \texttt{D}_{3}}$ are the Dirac masses. The constants $M_{1},M_{2},M_{3},m_{_{1}},m_{_{2}},m_{_{3}}$ are
unknown so the neutrino masses can not be calculated directly.
The case in which the mass matrix has the most general form involves so many parameters and becomes so complicated such that it cannot be solved therefore, we take a simpler form.

There is no exact data available about the neutrino masses but the cosmological measurements
\cite{k3} give a boundary of active neutrino masses
\begin{equation}
\sum_{i}m_{i}<0.17\rm{eV}.
\end{equation}
Also the neutrino mass differences \cite{k4} are given by experimental
measurements of solar, atmospheric, accelerator and reactor
neutrinos.
\begin{eqnarray}
\vert{\Delta m_{21}^{2}}\vert=(7.53\pm 0.18)\cdot 10^{-5} \rm{eV}^{2}, \nonumber \\
\vert{\Delta m_{32}^{2}}\vert=(2.44\pm 0.06)\cdot 10^{-3} \rm{eV}^{2}.
\label{10}
\end{eqnarray}
If we denote neutrino masses as $m_{\nu_{1}}$, $m_{\nu_{2}}$, and
$m_{\nu_{3}}$ then with the help of Eq.(\ref{10}) we can write
\begin{eqnarray}
\label{neutrinomass}
&~&\vert m_{\nu_{1}}^{2}-m_{\nu_{2}}^{2} \vert=\vert{\Delta m_{21}^{2}}\vert, \nonumber \\
&~&\vert m_{\nu_{3}}^{2}-m_{\nu_{2}}^{2} \vert=\vert\Delta m_{32}^{2}\vert.
\end{eqnarray}
 Putting the values of $\vert{\Delta m_{21}^{2}}\vert$ and$\quad
\vert{\Delta m_{32}^{2}}\vert$ in Eq.(\ref{neutrinomass}) and considering Mikheyev Smirnov Wolfenstein \cite{Wolfen,Mikhey} matter effects on solar neutrinos, we can get the following two sets of analytical formulas,
\begin{eqnarray}
&~& m_{\nu_{2}}^{2} = m_{\nu_{1}}^{2}   + 7.53 \times 10^{-5} \rm{eV}^2, \nonumber \\
&~& m_{\nu_{3}}^{2} = m_{\nu_{2}}^{2} + 2.44 \times 10^{-3} \rm{eV}^2.
\label{analytical}
\end{eqnarray}
and
\begin{eqnarray}
&~& m_{\nu_{2}}^{2} = m_{\nu_{1}}^{2}   + 7.53 \times 10^{-5} \rm{eV}^2, \nonumber \\
&~& m_{\nu_{3}}^{2} = m_{\nu_{2}}^{2} - 2.44 \times 10^{-3} \rm{eV}^2.
\label{analytical2}
\end{eqnarray}

Following the Koide's formula for leptons, we can write a relation for neutrinos as
\begin{eqnarray}
k_{\nu_{L}}^{2} = \frac{(m_{\nu_{1}}+m_{\nu_{2}}+m_{\nu_{3}})} {( \sqrt{m_{\nu_{1}}}+\sqrt{m_{\nu_{2}}}+\sqrt{m_{\nu_{3}}})^{2}},
\label{kvLform}
\end{eqnarray}
Using Eq.(\ref{analytical}), Eq. (\ref{kvLform}) can be rewritten as

\bea{}\label{kvL}
k_{\nu_{L}}^{2} = \frac{(m_{\nu_{1}}+ \sqrt{m_{\nu_{1}}^{2}+ 7.53 \times 10^{-5}\rm{eV}^{2}}+ \sqrt{m_{\nu_{1}}^{2} + 251.53 \times 10^{-5}\rm{eV}^{2}})}{( \sqrt{m_{\nu_{1}}}+ (m_{\nu_{1}}^{2}+ 7.53\times10^{-5}\rm{eV}^{2})^{1/4}+(m_{\nu_{1}}^{2} + 251.53 \times 10^{-5}\rm{eV}^{2})^{1/4})^{2}}
\eea

and using Eq.(\ref{analytical2}), as

\bea{}\label{kvLL}
k_{\nu_{L}}^{2} = \frac{(m_{\nu_{1}}+ \sqrt{m_{\nu_{1}}^{2}+ 7.53 \times 10^{-5}\rm{eV}^{2}}+ \sqrt{m_{\nu_{1}}^{2} - 236.47 \times 10^{-5}\rm{eV}^{2}})}{( \sqrt{m_{\nu_{1}}}+ (m_{\nu_{1}}^{2}+ 7.53\times10^{-5}\rm{eV}^{2})^{1/4}+(m_{\nu_{1}}^{2} - 236.47 \times 10^{-5}\rm{eV}^{2})^{1/4})^{2}}
\eea

The above equations can be solved to find the value of $ m_{\nu_{1}}$ if we can somehow constrain the value of $k_{\nu_{L}}^{2}$.

\section{Meaning of Foot's Geometrical Interpretation}\label{sec:last}

In this section we present Foot's geometrical
interpretation explaining what the vector
$\overrightarrow{u}=\left({1},\quad {1},\quad {1}\right)$ means.

The lepton masses have an equal status in Koide's relation which indicates a presence of some underneath symmetry.
With the help of this symmetry we can give a Koide's-like relation for Dirac neutrino mass terms.

The neutrino mass matrix is given by Eq.(\ref{mat}). We consider that there exists a symmetry such that the Dirac mass term of the three flavored neutrinos gives invariable result for the three generations of neutrinos which would mean
that, in the original neutrino mass matrix, three generation flavored
neutrinos have the same mass coefficient, that is,

\begin{equation}
m_{ \scriptsize\texttt{D}_{1}}= m_{\scriptsize\texttt{D}_{2}}= m_{\scriptsize\texttt{D}_{3}},
 \label{equal}
 \end{equation}
we can write a vector
\begin{equation}
\overrightarrow{U}=
\left(
\sqrt{m_{\scriptsize \texttt{D}_{1}}}, \quad \sqrt{m_{\scriptsize \texttt{D}_{2}}}, \quad \sqrt{m_{\scriptsize \texttt{D}_{3}}}
\right) ,\label{2ndlastsection}
\end{equation}
having characteristic
\begin{equation}
\frac{\overrightarrow{U}}{\vert \overrightarrow{U}\vert}=\frac{\left({1},\quad {1}, \quad {1} \right)}{\vert \left({1},\quad {1}, \quad {1} \right) \vert},
\label{3rdlastsection}
\end{equation}
which appears in Eq.(2) of Foot's paper \cite{Foot}. The lepton masses can form a vector
$\overrightarrow{V}=\left( \sqrt{m_{e}}, \quad \sqrt{m_{\mu}}, \quad
\sqrt{m_{\tau}} \right)$. The angle between
$\overrightarrow{U}$ and $\overrightarrow{V}$ is

\bea{}
\hspace{-9mm} \cos\theta =\frac{\left( \sqrt{m_{e}}, \quad \sqrt{m_{\mu}}, \quad
\sqrt{m_{\tau}} \right) \left(
\sqrt{m_{\scriptsize \texttt{D}_{1}}}, \quad \sqrt{m_{\scriptsize \texttt{D}_{2}}}, \quad \sqrt{m_{\scriptsize \texttt{D}_{3}}}
\right) }{\vert \left( \sqrt{m_{e}}, \quad \sqrt{m_{\mu}}, \quad
\sqrt{m_{\tau}} \right)\vert \vert \left(
\sqrt{m_{\scriptsize \texttt{D}_{1}}}, \quad \sqrt{m_{\scriptsize \texttt{D}_{2}}}, \quad \sqrt{m_{\scriptsize \texttt{D}_{3}}}
\right) \vert}
\eea
\begin{eqnarray}
\nonumber \\
&=&\frac{\left( \sqrt{m_{e}}, \quad \sqrt{m_{\mu}}, \quad
\sqrt{m_{\tau}} \right) \left(1, \quad 1, \quad 1
\right)}{\vert \left( \sqrt{m_{e}}, \quad \sqrt{m_{\mu}}, \quad
\sqrt{m_{\tau}} \right)\vert \vert \left(1, \quad 1, \quad 1 \right)\vert} \nonumber \\
&=&\frac{1}{\sqrt{3}}\frac{\sqrt{m_{e}}+\sqrt{m_{\mu}}+\sqrt{m_{\tau}}}{\sqrt{m_{e}+m_{\mu}+m_{\tau}}}.
\label{4thlastsection}
\end{eqnarray}
Using Eq.(\ref{koideee}), Eq.(\ref{4thlastsection}) gives
$\cos\theta=\frac{\sqrt{2}}{2}$, making $\theta=\frac{\pi}{4}$. This
relation can be expressed by vectors, as given by Fig[1].
\begin{center}
\begin{picture}(110,100)
\thicklines \put(55,10){\vector(0,1){60}} \put(55,65){\line(0,1){25}}
\put(55,10){\vector(1,1){60}} \thinlines
\put(35,40){$\overrightarrow{U}$}
\put(83,47){$\overrightarrow{V}$}
\put(20,0.){$\displaystyle $}\thinlines
\hspace{-18mm}Fig 1. The vectors $\overrightarrow{U}$ and $\overrightarrow{V}$ forms an angle $\frac{\pi}{4}$ .
\end{picture}
\end{center}
\vspace{1cm}
For Dirac neutrino mass terms, using Eq.(\ref{4thlastsection}), we can get
\bea{}\label{17}
\hspace{-1mm}
m_{\scriptsize \texttt{D}_{1}}+m_{\scriptsize \texttt{D}_{2}}+m_{\scriptsize \texttt{D}_{3}}= \frac{1}{3}(\sqrt{m_{\scriptsize \texttt{D}_{1}}}+\sqrt{m_{\scriptsize \texttt{D}_{2}}}+\sqrt{m_{\scriptsize \texttt{D}_{3}}})^{2}.
\eea
Because we have
\begin{eqnarray}
k_{l}^{2}=\frac{m_e+m_{\mu}+m_{\tau}}{(\sqrt{m_e}+\sqrt{m_{\mu}}+\sqrt{m_{\tau}})^2}, \nonumber \\
k_{\nu}^{2}=\frac{m_{\scriptsize \texttt{D}_{1}}+m_{\scriptsize \texttt{D}_{2}}+m_{\scriptsize \texttt{D}_{3}}}{(\sqrt{m_{\scriptsize \texttt{D}_{1}}}+\sqrt{m_{\scriptsize \texttt{D}_{2}}}+\sqrt{m_{\scriptsize \texttt{D}_{3}}})^2},
\label{5thlastsection}
\end{eqnarray}
we can write a new symmetric relation
\begin{equation}
k_{l}^{2}+k_{\nu}^{2}=1,
\label{6thlastsection}
\end{equation}
which is similar to
\begin{equation}
{\sin}^2\alpha+{\cos}^2\alpha=1.
\label{lastlastsection}
\end{equation}
The possible range of the coefficient $k_l^2$ in Koide relation is (1/3, 1). When three masses are identical (democratic), we have $k_l^2=1/3$; if three masses are strongly hierarchical, then $k_l^2=1$. $k_l^2=2/3$ is just the mean value of these two limits. Considering the degeneracy of the Dirac neutrino masses ($k^2_\nu = 1/3$, Eq.(\ref{equal})) will lead us to the above mentioned symmetric relations.


\section{Analytical Formula For Right Handed Neutrino Masses}
We know the matrix given in Eq.(\ref{matrix}) has 6 eigenvalues, when  $ M_i \gg m_{\scriptsize \texttt{D}_{i}}  (i=1,2,3)$ three would be given by Eqs.(\ref{eigenvalue1}-\ref{6}) and the rest of them would be equal to $ M_{i}  (i = 1 , 2 ,3)$. There is another way to find this second set of eigenvalues i.e. by using Eq.(\ref{6}), approximately similar relations can be found in \cite{new} and others.

The Dirac neutrino masses satisfy Eq.(\ref{17}) and also as discussed in Section (\ref{sec:last}), Eq.(\ref{equal}), the Dirac mass term gives invariable result for three generations of neutrinos. We can have Dirac neutrino masses to be proportional to the electroweak scale i.e. $\lambda_{EW} \approx 246 \rm GEV$. According to the SM gauge symmetries the right handed neutrino spinor in Eq.(\ref{3}) is uncharged indicating $M_i$ for $i= 1,2,3 $ a free parameter. While the Dirac masses are forbidden by electroweak gauge symmetry and can appear only after spontaneously breaking down through Higgs mechanism, as in the case of charged leptons, which employs Dirac masses naturally to be of the order of vacuum expectation value of Higgs field in SM, which is $ v = 246 GeV$ then $ v / \sqrt{2} \approx 174 GeV $ . If we consider Dirac neutrino particle masses to be of the order of electroweak scale, knowing the masses of neutrinos, we can get the masses for Majorana neutrinos, which is to say using expression in Eq(\ref{6}), right handed neutrino masses can be written as

\begin{equation}
M_{1} =\frac{m_{\scriptsize \texttt{D}_{1}}^2}{m_{\nu_{1}}}, \quad
M_{2} =\frac{m_{\scriptsize \texttt{D}_{2}}^2}{m_{\nu_{2}}}, \quad
M_{3} =\frac{m_{\scriptsize \texttt{D}_{3}}^2}{m_{\nu_{3}}}. \quad
 \label{Mmass}
\end{equation}
which also employs that no left handed neutrino should have a zero mass.

\section{Relation Between Left And Right Handed Neutrinos Masses}
Eq.(\ref{6thlastsection}) gives a relation between leptons and Dirac neutrinos masses. A similar kind of relation must exist for
the right and left handed neutrino masses, since left and right handed neutrinos take part in SeeSaw mechanism and follows Eq.(\ref{Mmass}), so

\begin{equation}
k_{\nu_{R}}^{2}+k_{\nu_{L}}^{2}=1,
\label{symmetryL&R}
\end{equation}

where $k_{\nu_{R}}^{2}$ would be
\begin{equation}
k_{\nu_{R}}^{2} = \frac{(M_{1}+M_{2}+M_{3})} {( \sqrt{M_{1}}+\sqrt{M_{2}}+\sqrt{M_{3}})^{2}},
\label{kvR}
\end{equation}
and $k_{\nu_{L}}^{2}$ is given by Eq(\ref{kvLform}).
Using Eq.(\ref{Mmass}), Eq.(\ref{kvR}) can be re-written as
\begin{equation}
k_{\nu_{R}}^{2} = \frac{( \frac{m_{\scriptsize \texttt{D}_{{1}}}^2}{m_{\nu_{1}}}+\frac{m_{\scriptsize \texttt{D}_{{2}}}^2}{m_{\nu_{2}}}+  \frac{m_{\scriptsize \texttt{D}_{{3}}}^2}{m_{\nu_{3}}})} {( \sqrt{\frac{m_{\scriptsize \texttt{D}_{{1}}}^2}{m_{\nu_{1}}}}+\sqrt{\frac{m_{\scriptsize \texttt{D}_{{2}}}^2}{m_{\nu_{2}}}}+\sqrt{\frac{m_{\scriptsize \texttt{D}_{{3}}}^2}{m_{\nu_{3}}}})^{2}}.
\label{kvRrrr}
\end{equation}
Since for neutrinos there are two possible mass schemes, the normal mass scheme in which $ m3 > m2 > m1$ or the inverted mass scheme, such that $ m2 > m1 > m3$. Both schemes should be considered to obtain the possible neutrino masses.

\subsection{Normal Hierarchy}
When the masses follow normal mass hierarchy i.e. $ m_{\nu_{1}}< m_{\nu_{2}}< m_{\nu_{3}}$.
\begin{equation}\label{extended}
= \frac{( \frac{m_{\scriptsize \texttt{D}_{{1}}}^2}{m_{\nu_{1}}}+\frac{m_{\scriptsize \texttt{D}_{{2}}}^2}{\sqrt{m_{\nu_{1}}^{2}+ 7.53 \times 10^{-5}\rm{eV}^{2}}} +
 \frac{m_{\scriptsize \texttt{D}_{{3}}}^2}{\sqrt{m_{\nu_{1}}^{2} + 251.53 \times 10^{-5}\rm{eV}^{2}}})} {( \sqrt{\frac{m_{\scriptsize \texttt{D}_{{1}}}^2}{m_{\nu_{1}}}}+ \sqrt{\frac{m_{\scriptsize \texttt{D}_{{2}}}^2}{\sqrt{m_{\nu_{1}}^{2}+ 7.53 \times 10^{-5}\rm{eV}^{2}}}}+  \sqrt{\frac{m_{\scriptsize \texttt{D}_{{3}}}^2}{\sqrt{m_{\nu_{1}}^{2} + 251.53 \times 10^{-5}\rm{eV}^{2}}}})^{2}} ,
\end{equation}

where we used Eq.(\ref{analytical}) to make Eq.(\ref{kvRrrr}) dependent only on one unknown parameter $m_{\nu_{1}}$.
Running an analysis on the value of $m_{\nu_{1}}$ using Eq.(\ref{kvL}), in such a way that $k_{\nu_{L}}^{2}$ gives a value, which when added to $k_{\nu_{R}}^{2}$ will satisfy Eq.(\ref{symmetryL&R}) completely. Since $k_{\nu_{R}}^{2}$ is dependent on the value of left handed neutrinos, we can obtain the left handed neutrino masses, i.e. $m_{\nu_{2}}$ and $m_{\nu_{3}}$ masses for the value of $m_{\nu_{1}}$, by using Eq.(\ref{analytical}). It is interesting to note that when we assume the above relation, in Eq.(\ref{symmetryL&R}), exist we can deduce the below mentioned important relations.
The only value which satisfies Eq.(\ref{symmetryL&R}) up to three decimal points gives left handed neutrino masses to be:

\begin{eqnarray}
\label{massvaluesLHN}
&~&m_{\nu_{1}}=0.00107\rm{eV},\nonumber \\
&~&m_{\nu_{2}}=0.00874\rm{eV},\nonumber \\
&~&m_{\nu_{3}}=0.05016\rm{eV}.
\end{eqnarray}
The above values of $\nu_{2}$ and $\nu_{3}$ neutrino masses are not only in accordance with the previous predictions \cite{Carl, Ma} up to three decimal points, but are also more precise. These masses follow normal mass hierarchy.

\begin{figure}[!h]
  \renewcommand\thefigure{2}
  \centering
  \includegraphics[width=13cm]{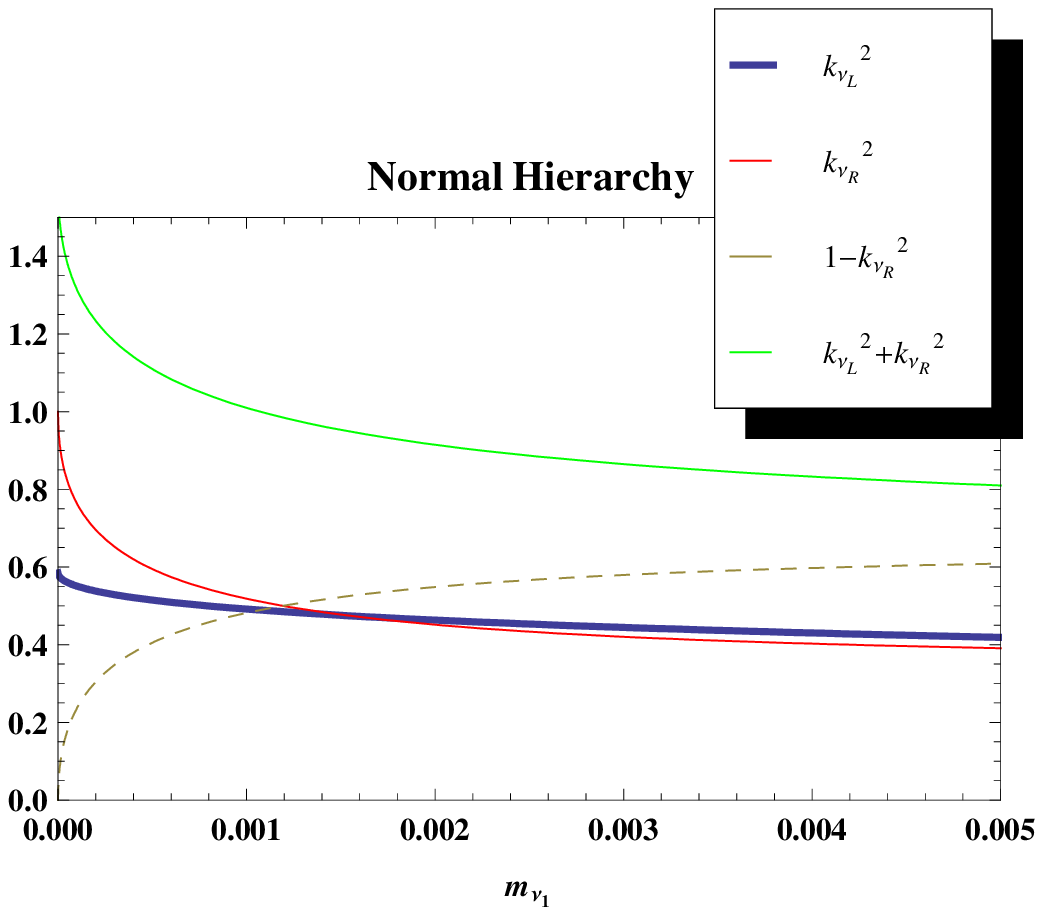}\\
  \caption{In case of normal hierarchy}\label{normal}
\end{figure}

Once we have obtained the left handed neutrino masses we can use it in Eq.(\ref{Mmass}) to obtain the right handed neutrino masses to be
 \begin{eqnarray}
 \label{valueR}
&~&M_1 = 2.8295\times10^{16} \rm{GeV},\nonumber \\
&~&M_2 = 3.4627\times10^{15} \rm{GeV},\nonumber \\
&~&M_3 = 6.0353\times10^{14} \rm{GeV}.
\end{eqnarray}
 which are approximately of $10^{16}$ GeV. In the case of normal mass hierarchy, the plot of Eq.(\ref{kvL}), Eq.(\ref{kvRrrr}) and Eq.(\ref{symmetryL&R}) dependent on $m_{{\nu}_1}$ is given in Fig.(\ref{normal}).

\subsection{Inverted Hierarchy}
  For the case of inverted hierarchy of left handed neutrinos masses, i.e. $  m_{\nu_{3}}< m_{\nu_{1}}< m_{\nu_{2}}$, we use Eq.(\ref{analytical2}) to make Eq.(\ref{kvRrrr}) dependent on only one unknown parameter $m_{\nu_{1}}$.

  \begin{equation}\label{extended2}
= \frac{( \frac{m_{\scriptsize \texttt{D}_{{1}}}^2}{m_{\nu_{1}}}+\frac{m_{\scriptsize \texttt{D}_{{2}}}^2}{\sqrt{m_{\nu_{1}}^{2}+ 7.53 \times 10^{-5}\rm{eV}^{2}}} +
 \frac{m_{\scriptsize \texttt{D}_{{3}}}^2}{\sqrt{m_{\nu_{1}}^{2} - 236.47 \times 10^{-5}\rm{eV}^{2}}})} {( \sqrt{\frac{m_{\scriptsize \texttt{D}_{{1}}}^2}{m_{\nu_{1}}}}+ \sqrt{\frac{m_{\scriptsize \texttt{D}_{{2}}}^2}{\sqrt{m_{\nu_{1}}^{2}+ 7.53 \times 10^{-5}\rm{eV}^{2}}}}+  \sqrt{\frac{m_{\scriptsize \texttt{D}_{{3}}}^2}{\sqrt{m_{\nu_{1}}^{2} - 236.47 \times 10^{-5}\rm{eV}^{2}}}})^{2}}.
\end{equation}

\begin{figure}[!h]
  \renewcommand\thefigure{3}
  \centering
  \includegraphics[width=13cm]{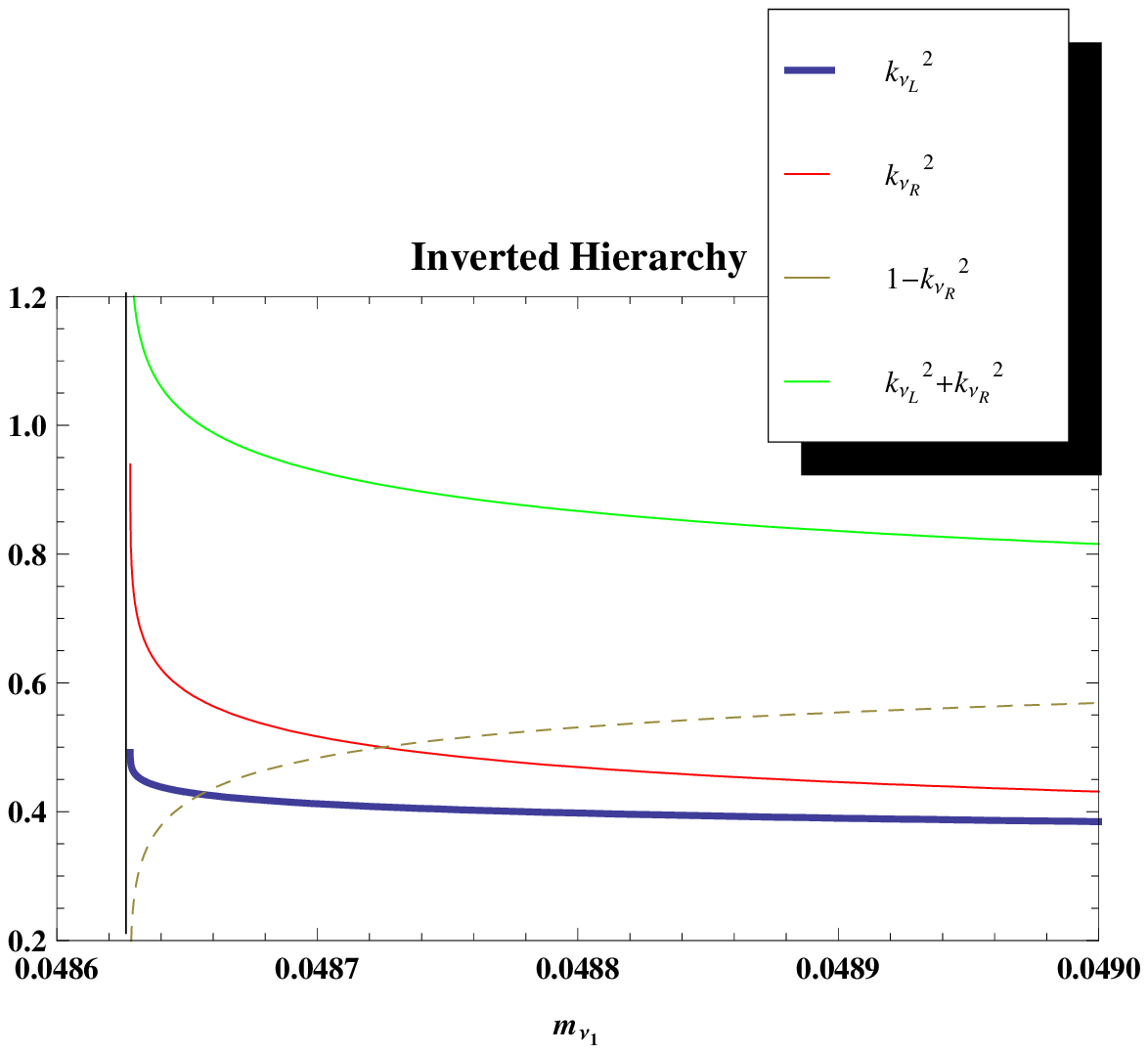}\\
  \caption{In case of inverted hierarchy}\label{hierarchal}
\end{figure}

After following similar steps and carefully looking for a solution in a very small range with $m_{{\nu}_1}\geq 0.04863$, Eq.(\ref{symmetryL&R}) satisfies giving
\begin{eqnarray}
&~&m_{\nu_{1}}=0.048655\rm{eV},\nonumber \\
&~&m_{\nu_{2}}=0.049423\rm{eV},\nonumber \\
&~&m_{\nu_{3}}=0.001625\rm{eV}.
\end{eqnarray}
and in turn masses of right handed neutrinos becomes
 \begin{eqnarray}
&~&M_1 = 6.22254\times10^{14} \rm{GeV},\nonumber \\
&~&M_2 = 6.12588\times10^{14} \rm{GeV},\nonumber \\
&~&M_3 = 1.86225\times10^{16} \rm{GeV}.
\end{eqnarray}
Fig.(\ref{hierarchal}) shows the plot of Eq.(\ref{kvL}), Eq.(\ref{kvRrrr}) and Eq.(\ref{symmetryL&R}) dependent on $m_{{\nu}_1}$ in the case of inverted mass hierarchy.

\section{Conclusion and Summary}
In this paper we showed how neutrino masses can satisfy Koide's relation. We discussed the Koide's mass relation and gave the Dirac mass terms a family symmetry. We consider the Dirac mass terms invariable and used this to give a meaning to the geometrical interpretation of Koide's formula given in \cite{Foot} which in turn leads to a new Koide's like relation for Dirac neutrino mass terms given by Eq.(\ref{17}). Koide's relation and this new Koide like relation for Dirac mass terms, if added together equals one which lead us to Eq.(\ref{lastlastsection}).
A similar kind of plane must exist for the left and right handed neutrinos because according to the seesaw mechanics the extreme masses of neutrinos are because of the interaction between them. This relation can solve the analytical formula for the left handed neutrinos in Eq.(\ref{kvL}) giving mass of three generations of neutrinos which are precise up to three decimal points with the previously proposed values. Also, we can find the masses of neutrinos following inverted mass hierarchy.

In our paper we define the Yukawa coupling to be 1 to take $m_{\scriptsize \texttt{D}}$ to be 174 GeV but we noticed that it does not affect the mass values given in Eq.(\ref{massvaluesLHN}) for the left handed-neutrino masses, even if changed over a wide range. Our model also proposes that;

\textbf{SeeSaw Mechanism}:
  Since mass of the left handed neutrinos are dependent on the mass of the Dirac neutrino and the right-handed neutrinos masses, the Koide like relation for the neutrino would be
  \begin{equation}
  {k_{L}}^{'2} = \frac{{k_{\nu}}^{2}}{{k_{\nu_{R}}}^{2}},
  \end{equation}
  where
  \begin{equation}
  k_{\nu_{R}}^{2}= \frac {1} {2},\label{kvr}
  \end{equation}
  so we can have,
  \begin{equation}
  {k_{L}}^{'2}= \frac {2} {3}.
  \end{equation}

  The above relation indicates that the SeeSaw mechanism may be the underlying mechanism and thus explains why neutrinos have such extreme masses. This relation gives the answer why neutrinos do not satisfy Koide's relation giving 2/3 as we have for leptons. The reason is that the neutrino masses are the ratio of Dirac and Majorana neutrinos so the ratio of the Koide's formula for these would give the same 2/3 as for leptons. This formula is to justify the 2/3 value and not to be used as the Koide's formula for neutrinos which gives the value approximately equal to 1/2 that can be calculated by using the mass values given in Eq.(\ref{massvaluesLHN}).

So far there is not much experimental evidence to explain the mechanism by which neutrinos gain mass and also with the present experimental energy range it is not possible to test the speculation about the existence of Majorana neutrinos but as mentioned above the values of neutrinos obtained in this paper are in accordance with the masses of neutrinos predicted by several other studies.\\

%
%
%
%

~~~\textbf{ACKNOWLEDGEMENTS} \\

The work is supported by National Natural Science Foundation of China (No. 11275017 , No. 11172008 and No.11173028). STI has the financial support from Chinese Scholarship Council. STI would like to thank Richard M. Woloshyn for his suggestions.
\\

%
%
%

\end{document}